# THE DISTANCE TO THE VIRGO CLUSTER FROM A RECALIBRATED TULLY-FISHER RELATION BASED ON HST CEPHEIDS AND A DEMONSTRATED TEERIKORPI CLUSTER INCOMPLETENESS BIAS

Allan Sandage

The Observatories of the Carnegie Institution of Washington, 813 Santa Barbara Street,

Pasadena, CA 91101, USA

and

G.A. Tammann

Astronomisches Institut der Universitat Basel, Venusstrasse 7, CH-4102 Binningen, Switzerland

G-A.TAMMANN@UNIBAS.CH

## ABSTRACT

The importance of the distance of the Virgo cluster in the ongoing debate on the value of the Hubble constant is reviewed. A new calibration of the Tully-Fisher 21-cm line width-absolute magnitude relation is made using Cepheid distances to 25 galaxies determined in various HST programs and reduced with the new Cepheid P-L relations that vary from galaxy-to galaxy. The calibration is applied to a complete sample of Virgo cluster spirals for the purpose of demonstrating the Teerikorpi cluster incompleteness bias. A diagnostic test is shown that should be useful in identifying the presence of bias in incompletely sampled data for distant clusters.

The bias-free TF distance modulus for the Virgo cluster is $m - M = 31.67$ ($D = 21.6$ Mpc). A systematic correction of 0.07 mag is made because the cluster members are redder in $B - I$ on average than the calibrators at a given line width, giving a final adopted modulus for the Virgo cluster core of $31.60 \pm 0.09$. If we assign a generous range of systematic error of ~ 0.3 mag, the distance D = 20.9 Mpc (m - M = 31.60) has a range from 24.0 Mpc to 18.2 Mpc (m - M between 31.9 and 31.3), and a Hubble constant of $H_o = 56$ between the limits of 49 and 65 when used with a cosmic expansion velocity of 1175 km s$^{-1}$ determined by the method of distance ratios of remote clusters to Virgo. This range overlaps our preferred value of $H_o = 62$ from the HST Cepheid calibration of type Ia supernovae recently determined. The TF modulus of Virgo determined here cannot be reconciled with the recent high value of $H_o = 72$ from Freedman et al.

Subject headings: Cepheids - Cosmology:distance scale- Galaxies:clusters:Virgo

## 1. INTRODUCTION

The value of the Hubble constant remains in contention even at the present conclusion of the two principal observational programs that use data from the Hubble Space Telescope (HST) for Cepheid variables in a series of calibrating galaxies. The high value of $H_o = 72 \pm 5$ km s$^{-1}$ Mpc$^{-1}$ (hereafter the units are assumed) from Freedman et al. (2001) defines a distance scale 16% shorter than derived by our consortium which uses type Ia supernovae as the distance indicator (Sandage et al. 2006) that gives $H_o = 62 \pm 5$. Although, predictably, commentators not in the arena themselves are often remarking that the error limits now overlap, this is not a good way to argue. One or the other (or both) studies contain systematic errors, and one, or both, are incorrect.

We have shown elsewhere (Saha et al. 2006) that the 16% difference between the distance scale of Freedman et al. and ours is not caused by significant differences in the basic photometry in the HST Cepheid fields, or in the treatment of the internal absorption of the Cepheids in the parent galaxies. The independent final photometry is remarkably consistent between the two programs. Our longer scale, and hence our smaller value of $H_o$, is due principally to a different slope and zero point for the Cepheid period-luminosity relation used by the two groups.

The Freedman consortium had originally adopted the same P-L relation that nearly all workers had used (including our group) until 2001. Had they kept with this relation, they would have obtained a value of $H_o = 65$, only 4% higher than ours. As explained by Freedman et al. (2001), they ultimately rejected this solution, presumably because it conflicted with the results of certain of their secondary indicators whose zero points they believed to be more reliable. Instead, they adopted the considerably more shallow slope to the P-L relation derived from new data by Udalaski et al. (1999) for the LMC Cepheids. This gave them fainter absolute Cepheid magnitudes and shorter distances to their program galaxies for their final report (Freedman et al. 2001, their Table 3).

The new development started in 2002. It was made by comparing separate P-L relations for the Galaxy and the LMC (using the new LMC Udalski data), showing that the P-L relation is not unique but varies in slope and zero point from galaxy-to-galaxy (Tammann & Reindl 2002; Tammann, Sandage, & Reindl 2003, sometimes TSR 03 hereafter; Sandage, Tammann, & Reindl 2004, sometimes STR 04 hereafter), likely due to metallicity differences. That the slopes and zero points of the P-L relation must be different between the Galaxy and the LMC is evident by the different temperatures and slopes of the edges of the Cepheid instability strip in these two galaxies (Sandage & Tammann 2006 for a review), requiring different P-L relations, as a function of metallicity. Support for this conclusion is from pulsation models by Fiorentino et al. (2002), and by Marconi, Musella, & Fiorentino (2005).

The consequences of the non-uniqueness of the P-L relation for the determination of Cepheid distances, and hence the Hubble constant, have been analyzed elsewhere (Saha et al.

2006, Sandage et al. 2006) and are not repeated here. It need only be said that our long distance scale with $H_o \sim 60$ results when the non-uniqueness of the P-L relation is taken into account and when a proper P-L relation is used for a relevant galaxy.

The purpose of the present paper is twofold. (1) We have used our HST Cepheid sample (Saha et al. 2006, Table A1) that defines our long distance scale to recalibrate the $M(B_T)^{o,i}$ blue absolute magnitude 21-cm Tully-Fisher relation and to determine therefrom a newly calibrated distance to the Virgo cluster core, defined as the mean of subclusters A and B, using a complete sample of Virgo cluster spirals. (2) Because the sample is as complete in spirals as is possible from the line-width data in the literature, it is used to demonstrate the Teerikorpi cluster incompleteness bias, again showing that one must sample > 4 magnitudes into the cluster luminosity function to obtain a correct value of the distance modulus to within 0.1 mag using the Tully-Fisher method in the B band for Virgo.

The plan of the paper is this. Section 2 reviews the proposition that the distance to the Virgo cluster can be used as a secure step with which to cross into the remote expansion field to determine the global value of the Hubble constant. Also in this section is a history of our 1974 correction to Hubble's distance scale by a factor of 5 to M101 at m - M = 29.3 (D = 7.2 Mpc) and by a factor of nearly 10 to the Virgo cluster at m - M = 31.7 (D = 21.9 Mpc). Section 3 cites a selection of a few of the key papers over the past 30 years that favored the short-to-intermediate distance scale with $H_o$ between 100 and $\sim$ 80, and those that favor the long scale with $H_o < 65$.

A new calibration of the Tully-Fisher relation is in section 4 using new HST Cepheid distances to 25 galaxies based on the use of different Cepheid P-L relations from galaxy-to-galaxy depending on metallicity (Saha et al. 2006, Table A1). Section 5 sets out a Tully-Fisher analysis of the distance to the Virgo cluster using the new calibration from section 4. The resulting data for the individual modulus of each galaxy in the sample is binned by line width and by apparent magnitude within each line-width interval in section 6 to illustrate the Teerikorpi cluster incompleteness bias in a more transparent way than we have done before. Section 7 uses the new TF distance to Virgo derived here of m - M = 31.6 to support the long distance scale with $H_o$ ~ 60.

## 2. THE IMPORTANCE OF THE VIRGO CLUSTER

The debate over the long and short distance scale has been in progress since the mid 1970s. It was early centered on the value of $H_o$ ~ 100 advocated by de Vaucouleurs (1982 for a review), compared with the long-scale value that gave $H_o$ ~ 55 when our first distance to the Virgo cluster and its relation to the global value of $H_o$ was published (Sandage & Tammann 1974b, 1975a,b, 1976 of the Steps series).

The Virgo distance was the last step in revising Hubble's highly compressed 1926-1953 distance scale where his distance moduli were 22.2 to M31, 24.0 to the NGC 2403/M81 group and also to M101, and 26.8 to the Virgo cluster. Our scale showed a progressive expansion of Hubble's scale, stretched by 2.2 mag at M31 that followed from Baade's 1950 value of m - M = 23.7 and revised by Swope to 24.4 as calibrated by Arp in Baade and Swope (1963), 3.6 mag for NGC 2403 at m - M = 27.56 (Tammann & Sandage 1968), 5.3 mag for M101 at m - M = 29.3 (Sandage & Tammann 1974a), and 4.9 mag at the Virgo cluster at m - M = 31.7 cited above.

These large stretching ratios to Hubble's scale are linear factors of 2.7 for M31, 5.2 for NGC 2403, 11.5 for M101, and 9.6 for the Virgo cluster. The factors are so large that they were generally discounted by the astronomical community at the time. The debate over our methods and results began then.

It is not useful to give here an extended history of the many facets of that debate over the past 30 years. Suffice to say, our 1968 value of the modulus of NGC 2403 at m - M = 27.6 has been confirmed to within a few tenths of a magnitude by modern observations of its Cepheids by Freedman & Madore (1988) using the French-Canadian CFHT 3.6-m telescope, overcoming the heavy criticism earlier by Madore (1976), and by Hanes (1982) in his rebarbative essay on many parts of our distance scale. In the fullness of time, most of these criticisms have proved to be incorrect. Our 1974-1975 distance scale is the same as the modern one to within the errors, generally to better than the 0.2 mag level.

Our stretched scale by a factor of ~ 10 to M101 with m - M = 29.3 compared with Hubble's 24.0 has also been confirmed from HST Cepheids by Kelson et al. (1996) and Kennicutt et al. (1998), although even as late as 1996 ours was the largest value of a number of independent estimates by others (see de Vaucouleurs 1982 and Kelson et al. for summaries). It turns out that each of other 10 determinations of the modulus discussed by de Vaucouleurs and again by Kelson et al. before HST contain systematic errors for a variety of reasons.

However, the distance to the Virgo cluster core remains in contention. If our modulus of ~ 31.6 is systematically correct, then the Hubble constant cannot be as large as the Freedman et al. value of 72. The conclusion is independent of any disagreement over putative local streaming motions and the infall velocity toward Virgo (actually the retarded expansion due to the Virgo-complex gravitational pull) because the global scale can be tied to Virgo by the <u>relative distances</u> of distant clusters to Virgo determined in a number of ways (Jerjen & Tammann 1993, Federspiel, Tammann, & Sandage 1998, hereafter FTS, and section 7 later here). This circumvents the need to know anything about the perturbations of the expansion velocity field by the local overdensity of the Virgo complex or the need to construct a model simulating it. The only datum needed to determine the global Hubble constant is the Virgo distance itself.

The importance of the Virgo cluster (and auxiliarily the Fornax cluster) in deciding between the two incompatible final summary $H_o$ results from HST is shown by Figure 1. This is the Hubble diagram plotted directly in distance vs. velocity. The diagram is taken from a conference summary report (Sandage, Tammann, & Saha 2001). It combines an original diagram by Freedman (1997, her Fig. 7) with our different distances and recession velocities to Virgo and Fornax. The Freedman distances in the diagram are 16.2 Mpc (m - M = 31.05) for Virgo and 18.2 Mpc (m - M = 31.3) for Fornax, requiring $H_o$ = 80 for Virgo using the abnormally high cosmic velocity of Virgo of 1300 km s$^{-1}$, and 1320 km s$^{-1}$, D = 16.2 Mpc and $H_o$ = 73 for Fornax. Here, Freedman uses the local perturbation model of Han and Mould (1990) to give these abnormally high values of the Virgo and Fornax velocities, corrected to the CWB kinematic frame.

Alternately, our moduli of 31.7 (D = 21.8 Mpc) shown for Virgo and 31.9 (D = 23.8 Mpc) with $v_{CMB}$ = 1175 for Virgo (cf. section 6) define a line with $H_o$ = 54, slightly to the right of the line marked 60. As said above, the Virgo free-expansion velocity that we use here is tied to the global expansion rate via relative distances to intermediate distant clusters by the method of Jerjen and Tammann (1993). The purpose of Figure 1 is to show that a definitive determination of the Virgo cluster distance promises a decision between the two incompatible distance scales and $H_o$ values.

One last vignette for this section; the problem with the Freedman scale (column 5 in Freedman et al. 2001, their Table 3) is even more stringent than suggested in Fig. 1 if their final Cepheid scale is used, based on their adoption of the shallow Udalski et al. P-L slope for the LMC. Instead of m - M = 31.05 for Virgo adopted by Freedman (1997) for Figure 1, rather their 2001 scale gives the even smaller Virgo distance modulus of m - M = 30.8 (D = 14.5 Mpc) based on the precept (Ferrarese et al. 2000) that the distance to the Virgo cluster core is given by the distances to the five Virgo cluster galaxies with Cepheids. These are NGC 4321, NGC 4496A, NGC 4535, NGC 4536, and NGC 4548, and where they have neglected, incorrectly we aver, NGC 4639 with the large modulus of (m - M)$_{Saha}$ = 32.20 as not belonging to the cluster. However, with a modulus of the cluster core at 30.8 instead of 31.05, and with $v_{cmb}$ = 1350 km s$^{-1}$ from Ferrarese et al., the Freedman et al. Hubble constant would be $H_o$ = 93. Alternatively, if $v_{cmb}$ = 1175 and m - M = 30.8, then $H_o$ = 81. These are 24% and 8% larger than that marked $H_o$ = 75 in Fig. 1, and larger still than $H_o$ = 72.

3. EARLY VALUES OF THE VIRGO CLUSTER DISTANCE BY VARIOUS AUTHORS

To gain some flavor for the early debate on $H_o$ and the Virgo cluster distance it is useful to recall a few key summary papers over the past several decades. These are mostly now of historical interest but they may be helpful to those seeking an entrance to the early literature and the origins of the debate. The listings that follow are not meant to be exhaustive. The text book by Rowan-Robinson (1984) is also useful here.

3.1 Results that favor the short distance scale

The summary by de Vaucouluers (1982) has already been mentioned where he discusses his many objections to our 1975 long scale beginning with his papers four years earlier (de Vaucouleurs 1978a,b, 1979). He concludes that the distance modulus of the Virgo cluster core is m - M = 30.37 (D = 11.9 Mpc) and that the global value of the Hubble constant is 95 $\pm$ 10.

An early challenge to our Virgo Cluster distance of 21.9 Mpc [(m - M) = 31.7] was by Tully and Fisher (1977) in their first application of the 21-cm distance method, calibrated with the distance modulus of M31 at m - M = 24.2, M33 at m - M = 24.6 and our moduli of 27.6 and 29.3 to the NGC 2403/M81 and M101 groups. Their resulting Virgo modulus was m - M = 30.6 $\pm$ 0.20 (D = 13.2 Mpc) giving, $H_o$ = 80 $\pm$ 8.

Pierce and Tully (1988), again using the 21-cm line width method but now with more calibrators, derived (m - M)$_{Virgo}$ = 30.88 $\pm$ 0.2 (D = 15.6 $\pm$ 1.5 Mpc) and $H_o$ = 85 $\pm$ 10.

Shanks et al. (1992), using the assumption that a single galaxy's distance as a member of the cluster defined the distance to the cluster, obtained m - M = 30.6 from the yellow giants in NGC 4523.

The most far reaching use of this assumption that the distance of a single galaxy can define the distance to the cluster core was by Pierce et al. (1994), claiming Cepheids in NGC 4571, and by Freedman et al. (1994) with their announcement of the actual discovery of Cepheids in NGC 4321 (M100). The Pierce et al. claim, as if Cepheids had been discovered, was that m - M = 30.9 $\pm$ 0.15 (D = 14.9 $\pm$ 1.2 Mpc) for NGC 4571, and therefore that $H_o$ = 87 $\pm$7 which they stated was well determined. The Freedman et al. claim of the discovery of Cepheids in M100 with HST was clearly correct but their assumption that M100 was in the core of the cluster was an assumption that could not be proved. Therefore their cluster modulus of 31.16 $\pm$ 0.2 was an assumption, and the derived $H_o$ 80 $\pm$ 17 was also an unproved assumption. Against the premise was the external evidence based on the ease of resolution of the brightest stars in both NGC 4571 and NGC 4321 (Sandage & Bedke 1994), putting these two galaxies on the near side of the cluster.

Jacoby, Ciardullo, and Ford (1990) derived a Virgo modulus of m - M = 30.74 $\pm$ 0.05, $H_o$ = 81 $\pm$ 6 from their claim that the luminosity function of planetary nebulae is universal and has a well defined vertical cut-off at bright absolute magnitudes. Tonry, Ajhar, & Luppino (1990) used their surface brightness fluctuation method on selected Virgo cluster E and S0 galaxies, calibrated using M31, to derive m - M = 31.15 $\pm$ 0.13, $H_o$ = 78 $\pm$ 6, for the cluster core.

A far ranging review by Jacoby et al. (1992) contains a comprehensive survey of the literature and the various methods to $H_o$ with the conclusion that the Virgo cluster modulus is 31.0 $\pm$ 0.23, and that Ho = 80 $\pm$ 11.

A summary paper by van den Bergh (1992) surveyed part of the same literature as Jacoby et al. In his Table 13 he lists the value of the Hubble constant derived by the principal players to that date, concluding that $H_o$ = 86 $\pm$ 6. In addition, he makes the claim that our distance scale starting in 1968 and continuing in the Steps series that began in 1974 contains a gross systematic error beginning at about m - M = 27 that vitiates our low value of $H_o$. Remarkably he does not discuss our distance to NGC 2403 with m - M = 27.6, or M101 with m - M = 29.3, both of which are correct. If he had plotted these distance moduli in his Figure 7 he would have concluded that our scale is correct and that the reason for the discontinuity he discusses, beginning at m - M = 27, lies elsewhere.

Freedman (1997, 1998) has also prepared conference summaries of her short distance scale. Two recent reviews updating her 1997 and 1998 summaries are her final 2001 NASA report and a conference review based on a decade of HST space science (Freedman et al. 2003).

### 3.2. Results that favor the long distance scale

There have also been many papers in the decade of the 1990s that favor the long distance scale with (m - M)$_{Virgo}$ near 31.6 and $H_o$ near 60. An early, highly significant, paper using the Tully-Fisher 21-cm method is by Kraan-Korteweg, Cameron, & Tammann (1988) with a strong calibration giving m - M = 31.60 $\pm$ 0.15, confirming our earlier result in Steps VII (Sandage & Tammann 1976) of m - M = 31.7 $\pm$ 0.08. KKCT give one of the earliest discussions of the Teerikorpi (1975, 1984; Bottinelli et al. 1986, 1987) incompleteness bias, which is also one of the main topics of the present paper (section 6 later here).

A method pioneered by van den Bergh and by Harris is to use the luminosity function (LF) of globular clusters, calibrated in the Galaxy using RR Lyrae star distances and integrated photometry of each cluster to obtain total magnitudes, and then to compare the magnitude at the maximum frequency of the LF in the Galaxy with that in program galaxies where the globular cluster magnitudes have been measured. Although the method has conceptual problems related to different formation histories of the globular cluster systems in different galaxies, nevertheless there is a literature on its application. Harris et al. (1991) observed the globular clusters in NGC 4365, NGC 4472, and NGC 4649, which, when added to those in M87 (van den Bergh, Pritchet, & Grillmair 1985; Grillmair, Pritchet, & van den Bergh 1986) gave a mean modulus for these Virgo cluster elliptical galaxies of m - M = 31.47 $\pm$ 0.25, $H_o$ ~ 70. They used an early version of the Galactic globular cluster luminosity function calibrated using an extant calibration of the absolute magnitude of Galactic RR Lyrae stars. A rediscussion gave m - M = 31.64 $\pm$ 0.25, $H_o$ = 57 $\pm$ 5 (Sandage & Tammann 1995). A total of eight E galaxies in Virgo with globular cluster luminosity functions eventually became available from sources summarized elsewhere (Tammann & Sandage 1999, Table 4) giving m - M = 31.70 $\pm$ 0.10, compared with five other methods that give <(m - M)> = 31.61 $\pm$ 0.09 (D = 21.0 $\pm$ 0.9 Mpc) for the Virgo cluster, giving the long distance scale of $H_o$ = 56 $\pm$ 3 (formal statistics).

A Tully-Fisher analysis of a complete sample of Virgo cluster spirals to B = 18, based on the Las Campanas Virgo cluster catalog (Binggeli, Sandage, & Tammann 1985) has been analyzed by Federspiel, Tammann, & Sandage (1998, hereafter FTS). HST Cepheid distances were used to 18 calibrating galaxies available to 1998 from various HST Cepheid programs.

From a detailed analysis, the Tully-Fisher cluster modulus based on 70 galaxies associated with the A and B subclusters centered in the Virgo complex, was m - M = 31.58 $\pm$ 0.24, $H_o$ = 57 $\pm$ 7.

Use of type 1a supernovae by Humuy et al. (1996) and by Phillips et al. (1999) each gave $H_o$ = 63 $\pm$ 5 with their version of the calibration of the mean SNe Ia luminosity at maximum. Tripp and Branch (1999) also derived $H_o$ = 63 $\pm$ 5 from their independent calibrations.

One could continue this summary covering perhaps 20 more determinations by others, both for the long and short distance scale. It is more useful to cite some of the recent comprehensive summaries of the debate that are scattered throughout the conference and review literature.

A review by Branch and Tammann (1992) of the usefulness of type Ia supernovae, both from the theory of the energy source at the explosion, and as a key element in the purely astronomical approach via the photometric distance scale ladder, gave $H_o$ = 57 $\pm$ 7. An update six years later (Branch 1998) strengthened the summary, giving $H_o$ = 60 $\pm$ 10.

A summary of the long distance scale was prepared by us for the 1997 Princeton 250th Birthday Celebration conference (Sandage, & Tammann 1997). Other reviews to 2003 have been made elsewhere (Tammann 1996, 1997, 1999; Sandage, Tammann, & Saha 1998, 2001; Tammann & Sandage 1999; Tammann et al. 2002; Tammann, Sandage, & Saha 2003). A summary of the 19th Texas Symposium of Relativistic Cosmology by Theuraeu and Tammann (2000) is also useful.

A series of recent analyses by Teerikorpi, Paturel, and collaborators (Teerikorpi & Paturel 2002; Paturel et al. 2002a,b; Paturel & Teerikorpi 2004) favor $H_o$ ~ 55 based on various bias corrections they have identified in works that have otherwise given the short distance scale of $H_o$ > 65.

Recent experiments on the Sunyaev-Zedovich effect and also the gravitational lensing favor the long distance scale.

A review for the Sunyaev-Zeldovich method to $H_o$ by Carlstrom, Holder, & Reese (2002, their Fig. 9) gives $H_o = 60 \pm 3$ for a cosmological model with omega (matter) = 0.3 and omega (cosmological constant) = 0.7. For a world model with an open geometry with omega (matter) = 0.3, omega (cosmological constant) = 0, then $H_o = 56$. For the unlikely model with omega (matter) = 1.0 requires $H_o = 54$. However, the values have a larger range of uncertainty when the possible systematic errors are considered.

Reporting a recent definitive measurement of the time delay of an optical outburst of the lensed quasar SBS 1520 + 530 by Burud, Hjorth Courbin et al. (2002), Magain (2005) summaries as: "These (lensing) values (of $H_o$) are in strong disagreement with the HST Key Project value of $H_o = 72$, but in agreement with $H_o = 58$ (from) Parodi et al. (2000) [from the HST Supernovae Calibration Consortium]." However, the statement appears to be too strong because it is sensitive to the model of the mass distribution of the lensing galaxy. Prasenjit Saha (private communication) tells us that the current data can be made consistant with $H_o$ anywhere between 60 and 80 at the 1 sigma level depending on the mass model of the lens.

## 4. A NEW CALIBRATION OF THE TULLY-FISHER LINE WIDTH-ABSOLUTE MAGNITUDE RELATION USING HST CEPHEIDS

### 4.1 The earlier calibration by Federspiel et al. (1998, FTS)

The calibration of the TF relation by Federspiel et al. (1998, FTS) used the distances determined from Cepheids with the HST available to late 1997. There were 18 Cepheid distances by several authors cited at the bottom of Table 2 of FTS. This system of distances had not been homogenized at that time as we have now done in the summary Paper IV of our consortium series (Saha et al. 2006). Rather, the FTS 1998 distances were determined from a P-L relation that was almost universally used then which was tied to the slope and zero point of the pre-Ulsaski LMC data adopted by Madore and Freedman (1991). This was before it was recognized that different P-L relations apply to different galaxies (TSR 03, STR 04).

Using these old adopted distance moduli (column 6 of Table 2 of FTS, 1998) for the 18 calibrators, plus the tabulated $B_T^{o,i}$ absorption corrected blue magnitudes from FTS column 3 of Table 2 and the log line width from column 12 of the same table as taken from the Lyon-Meudon Extragalactic Data Base (LEDA), gave the calibration equation (FTS equation 3),

$$M_{B(T)}^{o,i}(\text{FTS}) = -(6.97 \pm 0.02) \log LW_{20}(\text{LEDA}) - (2.35 \pm 0.05) \qquad (1)$$

It is essential here to emphasize which system of totally corrected magnitudes, $M_{B(T)}^{o,i}$, has been used for this equation.

It was stated by FTS in their descriptions of columns 3 and 4 of their Tables 2 and 3 that both the $B_T$ total magnitudes as uncorrected for internal and Galactic absorption, and the absorption $A_i$ and $A_g$ values used by FTS are from the RC3 (de Vaucouleurs et al. 1991). This is partially true but not entirely. The uncorrected (for Galactic and internal absorption) $B_T$ magnitudes, that form the base of the magnitude system used by FTS, is from the RC3. However, the correction for foreground Galactic absorption is not from the RC3 but is from Schlegel et al. (1998). These differ slightly from the values listed in the RC3. Hence, the FTS magnitude system, although close to the RC3 $B_T^{o,i}$ values, differ slightly from it. We use the FTS magnitudes for the remainder of this paper.

Also note again, following FTS (1998), that we use the RC3 $A_i$ internal absorptions, not those in the Revised Shapley-Ames Catalog (Sandage & Tammann 1987, the RSA), even as the RSA values have been used in Sandage et al. (2006) for reasons connected with field galaxies compared with cluster galaxies (Federspiel 1999, thesis). The $A^i$(RC3) differ from $A^i$(RSA) such that there is an offset of ~ 0.2 mag between the listed $B_T^{o,i}$(RC3) and the $B_T^{o,i}$(RSA) totally corrected magnitude systems (depending slightly on galaxy type), and hence between $B_T^{o,i}$(FTS) used here and $B_T^{o,i}$(RSA).[1]

The consequence for the determination of the Virgo cluster distance modulus is nil provided that we use the same magnitude system for the calibrators (Table 1 set out later here) as for the program galaxies (Table 2 later here), as was done in Tables 2 and 3 of FTS.

In updating equation (1) from FTS (1998), 11 more Cepheid calibrator galaxies became available to Federspiel (1999) than we used in FTS. This gave Federspiel (PhD thesis, Table 2.1, p 14) a value of -2.42 for the zero point in eq. (1) instead of -2.35.

However, because of the major change made in 2003 by our new understanding of the non-uniqueness of the P-L relation (Tammann, Sandage, & Saha 2003; TSR 03; STR 04, Sandage et al. 2006), the zero point of the calibration equation must again be reconsidered using our new system of Cepheid distances based on the P-L relations that vary from galaxy-to-galaxy. The main purpose of this paper is to use our new data base of Cepheid distances from Saha et al. (2006, Table A1) to recalibrate the TF relation, and to apply the new calibration to the Virgo TF data to see the effect on the derived Tully-Fisher cluster modulus to the A plus B cluster core.

4.2. A new calibration using the Cepheid distances in Saha et al. (2006, Table A1)

W set out the new calibration data in Table 1. Column 1 names the galaxy whose Cepheid distance has been determined on our 2006 distance scale. Column 2 lists the galaxy type and luminosity class taken from the 1987 edition of the RSA. Column 3 shows the luminosity class, coded as L = 1 for class I, 2 for I-II, 3 for II, 4 for II-III, and 5 for III, and 6 for III-IV. Column 4 gives the new distance modulus on our 2006 scale, taken from Saha et al. (2006, Table A1) and marked Saha 06. Column 5 shows the fully corrected apparent magnitude from TFS. Column 6 is the fully corrected absolute magnitude found by subtracting column 4 from column 5. Column 7 is the log line width from the Lyon-Meudon Extragalactic Data Base (LEDA) data base read at the 20% level and corrected for inclination as listed by FTS. We use the LEDA line widths throughout this paper[2].

---

[1] Comparisons of the listed $B_T^{o,i}$ magnitudes in FTS for 90 overlaps with the RC3 values listed in the RC3 show that the RC3 totally corrected magnitudes, $B_T^{o,i}$, are systematically $0.04 \pm 0.01$ (rms = 0.13) mag fainter than the fully corrected magnitudes in FTS used here. But, as emphasized above, this difference has no consequence for our derived distance modulus because we use the same magnitude system for the calibrators as for the program galaxies. Differences with the RC3 cancel in the mean.

Fixing the slope at -6.97 taken from FTS, and using the distance moduli and the resulting absolute magnitudes of the calibrators and the line widths from Table 1, gives the calibration of,

$$M_{B(T)}^{o,i}(\text{FTS system}) = -6.97 \log LW_{20}(\text{LEDA}) - (2.49 \pm 0.10). \qquad (2)$$

The zero point of -2.49 is larger than -2.42 from Federspiel's thesis and larger than the value of -2.35 from FTS that used only 18 calibrators; our new 2006 Cepheid distance scale from Saha et al. (2006) is longer than that used by FTS by $0.059 \pm 0.02$ (rms = 0.119) mag[3]. Said differently, the Cepheid scale used by FTS before our 2006 Cepheid revision is shorter than our final 2006 scale from Saha et al. (2006, Table A1) by $0.06 \pm 0.02$ mag, giving the brighter zero point in equation (2) of -2.49.

---

[2] We note that these LEDA line widths average 0.033 dex smaller than what can be calculated from the RC3. Hence, caution is to be used in applying the calibration to be made here using the LEDA calibration equation to other systems of line width. The large random errors and the systematic differences between line-width systems is one of the major uncertainties in using the Tully-Fisher method because the slope of the TF relation is so steep (7.0 in B rising to 11.0 in H).

[3] It is of interest to compare the 2006 Cepheid scale used here with the old (pre 2001) and new (post 2001) Cepheid scales of Freedman et al. (2001, their Table 3), the new being the scale that provides the root from which Freedman's et al. value of $H_0 = 72$ stems. The old and new Freedman et al. scales are set out in columns 2 and 5 of Table 3 of Freedman et al. (2001). Comparison shows that our distances are $0.097 \pm 0.030$ mag (rms = .141 mag) more distant than the Freedman et al. old scale (neglects NGC 5253 and NGC 3627), and $0.224 \pm 0.033$ mag (rms = 0.170 mag) (neglects NGC 5253 and NGC 3627) more distant than the Freedman new scale (their col. 5). Our distance moduli are larger by these amounts than Freedman's, and hence our absolute Cepheids are brighter by the same amount, leading to our smaller Hubble constant.

The data for the 25 calibrators from Table 1 are shown in Figure 2, where the ridge line is equation (2). The two envelope lines are arbitrarily put at one magnitude brighter and fainter than the ridge line.

## 5. THE VIRGO CLUSTER SAMPLE BINNED BY LINE WIDTH AND APPARENT MAGNITUDE

Our cluster sample is the same as was used by FTS (1998) in their Table 3. All galaxies that were considered to be cluster members by Bingelli, Sandage, Tammann (1985, BST) (or in a few cases added by others), based on morphology alone as classified from the DuPont telescope from the Las Companas Virgo survey, were first considered. The fiducial sample chosen by FTS have inclinations greater than $45^o$ and are members of either subcluster A or B as defined by Binggeli, Tammann, & Sandage (1987) as refined by Binggeli, Popescu & Tammann (1993). We have added to this fiducial sample all other galaxies with inclinations between $30^o$ and $45^o$ because of the insensitivity of the final result on inclinations that are as small as $30^o$ (FTS).

We have expanded the fiducial sample to include the 52 galaxies in the last two sections of Table 3 of FTS that are outside subclusters A and B but nevertheless were considered to be members of the wider Virgo complex by BST (1985). Not all galaxies in this added group were kept, because some deviated greatly from the resulting TF correlation and are clearly

---

Had we used either of the two distance scales of Freedman et al (2001, their Table 3, columns 2 and 5) the zero point of equation (2) would have been either $- 2.369 \pm 0.096$ (rms = 0.478), or $- 2.230 \pm 0.095$ (rms = 0.476).

background. They can be identified by comparing the "outsider" list in Table 2 with the last two sections of Table 3 of FTS. More on the final data base selection below.

For the reasons to be apparent in section 6, the data are binned by line width, and within each line width interval are ordered by apparent magnitude as set out in Table 2. Column 1 is the name; column 2 is the type taken from the RSA if it exists or from the VCC; column 3 is the coded luminosity class on the system of Table 2; column 4 is the log line width corrected for inclination on the LEDA system taken from Table 3, col. 11, of FTS; column 5 is the fully corrected apparent $B_T^{o,i}$ magnitude taken from col. 4, Table 3, of FTS; column 6 is the resulting distance modulus calculated as if each galaxy has the ridge-line absolute magnitude from equation (2) on the distance scale of Saha et al. (2006, Table A1); column 7 is the subcluster or other region of the Virgo complex in the notation used by FTS. In the non-subcluster regions the added galaxies are assigned to either subcluster regions on the basis of declination; the subcluster "A" region have declinations greater than $10^o$; the subcluster "B" region have declinations less than $10^o$ as in the diagram in BPT (1993).

Figure 3 shows the Tully-Fisher correlation using the data in Table 2 for the subclusters A and B alone in the upper panel, and for the A + B subclusters plus the added sample on the outside of these two subgroups. The ridge line in each panel is calculated with a modulus of m - M = 31.67 determined in the next section. The envelope lines are again placed one magnitude brighter and fainter than the ridge line.

In the added "outside" sample, many of the additions defined the same ridge line shown in Fig. 3, but a subgroup showed a fainter, yet well defined correlation of magnitude with line width that has the same slope as that shown in Fig. 3. This group is clearly in the background. It was removed from an earlier version of Table 2 on the basis of being outside the lower envelope in the bottom panel of Figure 3.

6. THE VIRGO DISTANCE AND A DEMONSTRATION OF THE TEERIKORPI CLUSTER INCOMPLETENESS BIAS

6.1 Demonstration of the bias in the distance modulus when the cluster luminosity function is incompletely sampled

The reason for binning the data into line-width intervals and then by apparent magnitude within each interval is evident from Figure 4. Based on this diagram we begin a new discussion of the Teerikorpi cluster incompleteness bias that is set out here in a different way than we did in Federspiel, Sandage, & Tammann (1994, their Figs. 5, 6 and 8) and/or in Sandage, Tammann, & Federspiel. (1995, their Figs. 2, 3, and 10), or in the way done by Giovanelli et al. (1997a,b; 1998; 1999).

Fig. 4 shows the ridge and envelope lines from Fig. 3 plus a magnitude-sampling limit line put at B = 11.5, which is 1.7 magnitudes into the luminosity function (the brightest galaxy in Table 2 has $B_T^{o,i}$ = 9.8). Consider three log line-width intervals from 2.70 to 2.75, 2.60 to 2.65, and 2.50 to 2.55 shown by the three vertical strips.

The ridge and envelope lines mean this. The ridge line is the locus of the calibration equation (2). To use the TF relation, equation (2) provides the ridge-line (most probable) absolute magnitude for a galaxy with a measured line width. Of course any given galaxy will not generally have exactly this absolute magnitude if there is intrinsic dispersion in the TF relation, or if there is a back-to-front distance range, or both. In the case of intrinsic dispersion there is a distribution of true absolute magnitudes at a given line width, which is the line-width specific luminosity function. For any given galaxy we do not know where in the luminosity distribution this true absolute magnitude lies, and this is the cause of the incompleteness bias, seen by the standard analysis that follows.

Consider the log line-width interval in Fig. 4 from log LW 2.70 to 2.75 for the right-hand strip. Suppose that all galaxies in the cluster are at the same distance (i.e. with zero back-to-front ratio). Galaxies near the upper envelope line in Figure 4 are those that are overluminous for their line width relative to the most probable value (on the ridge line). However, applying the fainter absolute magnitude calibration of the ridge line to such intrinsically bright galaxies gives too small an m - M distance modulus for them. Next consider galaxies near the lower envelope, analyzed in the same way. Here, all distance moduli derived from galaxies below the ridge line will be calculated to have too large distance moduli. However, if the strip is filled symmetrically (equal numbers above and below the ridge line) the mean modulus found by averaging the individual estimated moduli of all galaxies (all of which are incorrect except for those on the ridge-line) within the strip will be the true modulus to within statistics.

However, if the strip is not sampled completely, as in the second and third strips in Figure 4, the average of the individual moduli of the galaxies in the accessible portions of the strips (i.e that are brighter than the limit magnitude at 11.5 in the diagram), will be progressively in error. The resulting average of the modulus values will then be too small by amounts that will decrease as the fraction of the accessible sample to a complete sample within the strip increases with fainter limit lines. This is the incompleteness bias. Averaging over all line widths when the magnitude-limit line reaches $B = 15$ will only then give the correct distance modulus for the cluster if the strips are filled symmetrically. A similar analysis applies for the case of a significant back-to-front ratio even in the case of zero intrinsic dispersion in the TF correlation itself.

Although the change of the calculated mean distance modulus as we proceed from bright to faint apparent magnitude cut-offs in any given vertical strip is large if the dispersion in magnitude at any given line width is large (amounting to the two magnitudes from the bright-to-faint envelope lines in the B band here for Virgo), a summing over all line widths dilutes the effect. This is because as the magnitude limit line moves faintward, a larger fraction of the vertical strips from right to left in Fig. 4 are bias free, and will begin to dominate when we sum over all line widths. Nevertheless, the bias persists even to B = 14 at the 0.1 mag level in the B band due even to a small incomplete sampling. Of course, the bias is smaller if the intrinsic dispersion is smaller, as in photometric bands at longer wavelengths. However, due to the bias, the <u>estimated</u> true dispersion from biased data is always smaller than the <u>true</u> dispersion, as shown below.

Of course the bias effect can be modeled by knowing the shape of the luminosity function and dispersion within each strip, together with the population ratios along the ridge line, but such information is not easily available. However, the Virgo cluster data themselves, binned as in Table 2 by line width and apparent magnitude, can be used to illustrate all aspects of the bias and their consequences as the sampling magnitude is varied from B = 10 to 15.

6.2 Demonstration of the bias using the Virgo cluster data from Table 2

Figure 5 shows how the calculated individual modulus values change progressively with apparent magnitude in each of the line-width intervals in Table 2. As discussed above, if each line-width interval were to be sampled completely, the mean moduli in each interval would be the true modulus to within statistics. However, for various apparent magnitude limits (vertical lines put at various values of the abscissa in Fig. 5), various line-width intervals are incompletely sampled. To guide the eye, a vertical line at B = 12 is put in Fig. 5.

The situation is shown in more detail in the tabulations of summations as functions of apparent magnitude shown in Tables 3-5.

The matrix Table 3 shows the mean calculated moduli <u>between</u> the listed magnitude intervals for individual line-width intervals calculated by summing the data in Table 2. Tabulated are the calculated mean distance moduli for each line-with interval between the listed magnitude limits, obtained by averaging the values in Fig. 5 between magnitude-limit lines drawn 0.5 mag apart. In parentheses are the numbers of galaxies making up the listed mean. The table illustrates the run of calculated <m - M> averages that would be obtained by averaging the run of the individual values in Fig. 5 within the stated line-width and magnitude intervals.

An important use of such a tabulation as Table 3 but applied to other clusters would be to check on the incompleteness of the samples. For instance, in the absence of bias, the <m - M> values averaged over all magnitude intervals for a given line width interval listed at the foot of each line-width column should be constant within statistics. The numbers here for Virgo conform to this requirement, showing the uniformity of the filling of the data between the envelope lines in Fig. 3. If the Virgo cluster sample here had been appreciably incomplete, these numbers would show a trend with line width, which they do not. To illustrate this central point, draw a horizontal line at any given magnitude level in Table 3, or a vertical line in Fig. 5 as we have done at $B = 12$, and see the effect of the completeness at every line width that is not complete to any given magnitude limit.

Of course, the large variation in <m - M> at any given line width interval as the magnitude limit is changed is diluted by summing over all line widths, seen in the data in the last column of Table 3. The total range of calculated mean moduli is now only about one magnitude, whereas it is almost two magnitudes at any given line-width interval seen in the body of the table.

Table 4 shows the result of averaging all moduli <u>up to</u> the listed magnitude limits in column 1 for each line-width interval of the total sample (subclusters A + B plus the "outsiders"). The entries are the sums of the distributions <u>between</u> the magnitude limits in Table 3. Column 7 is what the observer will see if the data are not separated by line width as the magnitude limit of the observations is made fainter.

From these Virgo cluster data, the derived bias-affected modulus varies between 31.38 and 31.67 as the depth of penetration into the cluster luminosity function is increased from B = 10 to 15 as seen in column 7. Also seen in column 8 is the increase in the apparent rms spread in the m - M distributions at each magnitude limit, ranging from rms = 0.37 mag to 0.59 mag as the sampling is made from B = 10 to 15. The observed scatter is always too small when bias is present, as has been emphasized elsewhere (STF 1995). A false impression of the true dispersion has sometimes been written in the literature.

Table 5 is the same as Table 4 but with the subclusters A and B treated separately to test if there is a significant modulus difference between them, as has often been suspected by others. There is evidence for and against a difference, as summarized by FTS (1998), by Ferrarese et al. (2000), and undoubtedly by others. Table 5 shows that the bias-free mean modulus of the subcluster A plus its outsiders (56 galaxies), is $<m - M> = 31.57 \pm 0.07$ (rms = 0.551 mag), and $<m - M> = 31.80 \pm 0.10$ (rms = 0.613 mag) for subcluster B plus its outsiders (39 galaxies). Hence, this analysis supports a modulus difference of $0.23 \pm 0.12$ mag. The overlap is large.

Figure 6, top panel, shows the decrease in the <m - M> bias as the magnitude limit is made fainter from B = 11 to 15. The bottom panel shows the increase in the observed rms scatter over the same magnitude interval. The data are from columns 7 and 8 of Table 4.

Figure 7 shows the histogram of the individual m - M values from Table 2 for the total sample (subclusters A and B plus the "outsiders"). The distribution is sensibly Gaussian with a mean of <m - M> = 31.67 which we aver to be bias free. It must be emphasized that the spread in Fig. 6 is not due only to a back-to-front distance spread ($\pm$ 0.19 mag for a $10^o$ diameter cluster projected in the sky), but is primarily due to the larger spread in distance moduli (artificial) caused by the spread of the limit line relative to the ridge-line in Figure 3.

The form of the diagnostic distributions of the individual calculated m - M values that are implicit in Tables 3-5, such as the systematic run of modulus values with apparent magnitude in Fig. 5 for different line-width intervals, or the large difference in the inferred modulus between apparent magnitude limits as a function of line width (Table 4), can be expected to test for incompleteness bias in other clusters where the sampling is not as complete as for the Virgo cluster here. Examples include the clusters used by Giovanelli et al. (1997a,b, 1998), Dale et al. (1999), and Sakai et al. (2000).

Of course, the test depends crucially on knowing the correct unbiased slope of the TF relation. This generally cannot be obtained from an incomplete cluster sample because such a slope itself is biased. The slope must be determined by some other means, as FTS did for Virgo using the cluster sample that is known to be complete. But once the correct unbiased slope to the TF relation in any given photometric band is determined, any systematic variation of the mean moduli in the separate line-width intervals, such as is set out at the foot of Table 3, will signal that an incompleteness bias is present.

# 7. STEPPING FROM THE VIRGO CLUSTER TO THE REMOTE EXPANSION FIELD TO OBTAIN $H_o$(GLOBAL): REMAINING PROBLEMS

The problem of circumventing the effect on the local expansion velocity field due to the overdensity of the Virgo complex, and therefore in deciding what velocity of the Virgo cluster to use in combination with its distance to determine $H_o$(global), was solved by Jerjen and Tammann (1993). Their generalization of using the distance <u>ratios</u> to Virgo of a number of distant clusters outside the local peturbation field decisively changed the reliability of the method by decreasing the effect of the random motion of any given cluster (such as when using only the Coma cluster, as was often done in the past) relative to the cosmic microwave background.

The relative Hubble diagram by Jerjen and Tammann, made from the distance ratios to Virgo, and of course putting the Virgo cluster into the diagram at a distance ratio of 1, has the equation (Fig. 14 and eq. 13 of FTS 1998),

$$\log v_{CMB} = 0.2[(m - M) - (m - M)_{Virgo}] + (3.070 \pm 0.011). \qquad (3)$$

Hence, the expansion velocity that the Virgo cluster would have had in the CMB kinematic frame in the absence of the local velocity perturbation is $v_{CMB}$(Virgo) = $1175 \pm 30$ km s$^{-1}$.

With our new Tully-Fisher distance from the last section of D = 21.58 Mpc (m - M = 31.67), $H_o$ = 54 This is 15% lower than our preferred value of 62 from the supernovae experiment, calibrated with Cepheid distances (Sandage et al. 2006). Is the difference significant?

## 7.1 The problem of systematic errors

Assigning a realistic error requires knowledge of the systematic uncertainties in the TF method itself such as the need to know accurately the internal absorption corrections both for the calibrators and for the program galaxies, the need to know accurate inclination corrections to the line widths, knowledge of the effect of internal random motions on the observed line widths, knowledge of whether the calibrators have the same mean parameters such as line-width inclination corrections and/or the same colors as the cluster galaxies, and a myriad of other problems associated with galaxy photometry. An underestimate of these systematic errors has always plagued the method.

One such systematic correction of 0.07 mag, which we adopt, reduces our final modulus to m - M = 31.60 $\pm$ 0.09 because the cluster members at a given line width are redder on average than the calibrators (FTS 98, section 8 there). If we again generously assign a range of systematic error as large as ~ 0.3 mag, then m - M ranges between 31.9 and 31.3, or D between 24.0 Mpc and 18.2 Mpc. Used with 1175 km s$^{-1}$ from Jerjen and Tammann (1993), gives $H_o$ = 56 with an outside range between 49 and 65, just accommodating our Cepheid supernovae value of 62 (Sandage et al. 2006).

But if we can just accommodate the TF value here compared with our preferred supernova value of 62, the Freedman consortium value of $H_o$ = 72 has a serious problem if they insist (Ferrarese et al. 2000) that the Virgo modulus is 30.8, based on the four Virgo galaxies with Cepheid distances (ignoring as they do NGC 4639). With their m - M = 30.8 (D = 14.5 Mpc) and $v_{CMB}$ = 1175 km s$^{-1}$, their Hubble constant via the route of the Virgo cluster (Freedman et al. 1994) is $H_o$(Freedman) = 81, not 72.

To obtain $H_o$ = 72 with $v_{CMB}$(Virgo) = 1175 km s$^{-1}$ requires that they adopt a Virgo distance of 16.3 Mpc, or m - M = 31.06, Either they must abandon their modulus of 30.8, or they must revise back to their old Cepheid distance scale (Freedman et al. 2002, Table 3, col 2), instead of their new scale (their Table 3, col. 5). They must also continue to ignore the distance to Virgo derived from the TF method itself, which on their new Cepheid scale would be 0.22 mag smaller than on our Cepheid scale (footnote 3), or 31.60 - 0.22 = 31.38.

However, they cannot tolerate this distance because it would give them $H_o = 62$ if they were to use the Jerjen/Tammann Virgo expansion velocity of 1175 km s$^{-1}$ in the CMB frame with m - M = 31.38. Of course, if they were to still maintain that the CMB Virgo velocity is 1350 km s$^{-1}$ as in Ferrarese, violating the Hubble diagram of Jerjen and Tammann, and if they were also to use m - M = 31.38 based on our modulus when it is made smaller by the 0.22 mag difference in our Cepheid scales (Saha 2006, Table A1 vs. Freedmann et al, 2001, Table 5, col.5), then their value of $H_o$ would in fact be be 72, but such a Virgo $v_{CMB}$ velocity of 1350 km s$^{-1}$ is unrealistic based on the determination by Jerjen and Tammann that is free from all local perturbation models.

However, if, for some reason, one does not wish to rely on the distance ratios to Virgo of Jergen & Tammann, but rather to rely on a perturbation model to derive the Virgo velocity relative to the CMB, the numbers are these; The observed mean heliocentric velocity of the Virgo core is $1050 \pm 35$ km s$^{-1}$ (Bingelli et al. 1993), which transforms to 932 km s$^{-1}$ relative to the centroid of the Local Group. Using v(infall) = 220 km s$^{-1}$ for the Virgo infall velocity gives $v_{Virgo}(CMB) = 1152 \pm 35$ km s$^{-1}$. This is the same as 1175 km s$^{-1}$ to within statistics.

Other determinations of a low value of the cosmic Virgo velocity exist, one example of which is Hamuy et al. (1996), but all have such large errors that they are not competitive with the values cited above.

We finally note that if Freedman et al. were to continue using a Virgo modulus of 30.8 and $v_{cmb}$ = 1350 km s$^{-1}$ then $H_o$ becomes 93; or with 1175 km s$^{-1}$, then $H_o = 81$, neither of which is 72. It may be for these reasons that Freedman et al. (2001) have ignored the TF distance to the Virgo cluster in their summary paper.

Clearly, there is still work to be done in reconciling the TF distance to Virgo derived here with the Freedman short Virgo distance modulus of m - M = 30.8 and hence of deciding between the often quoted value of $H_o$ = 72 and our long scale with $H_o$ ~ 60 (Sandage et al. 2006), based on type 1a supernovae analyzed with a Cepheid P-L relation that varies from galaxy-to-galaxy.

FIGURE CAPTIONS

Fig. 1. The Virgo and Fornax clusters using m - M = 31.7 (D = 21.9 Mpc) for Virgo and m - M = 31.9 (D = 24.0 Mpc) for Fornax, and using 1175 km s$^{-1}$ for the fully corrected Virgo cosmic velocity, are superposed on a diagram by Freedman (1997) showing $H_o$ = 75 where her assumed modulus to Virgo is 31.05 (D = 16.2 Mpc), and 31.3 (D= 18.2 Mpc) for Fornax (before her 2001 Cepheid correction to an even shorter distance scale), and using the abnormally high values of the Virgo cluster velocity near 1350 km s$^{-1}$, as tied to the CMB by Han and Mould (1993). Had Freedman used her later adopted modulus of 30.8 for Virgo, retaining $v_{cmb}$ = 1350 km s$^{-1}$, she would have concluded that $H_o$ ~ 93 rather than 75 in this diagram.

Fig. 2. The calibration of the blue Tully-Fisher relation from the data in Table 1 based on the Cepheid distances of 25 calibrating galaxies, on the distance scale listed by Saha et al. (2006, Table A1). The ridge line is equation (2) of the text. Envelope lines are drawn one mag brighter and fainter than the ridge line. The magnitude system is that of FTS (1998) which is close to but not exactly that of the $B_T^{o,i}$ system listed in the RC3 (footnote 1).

Fig. 3. The correlation of line width and apparent magnitude from the data in Table 3 for the cluster sample. Top panel is for the fiducial sample of FTS for the subclusters A and B plus galaxies with inclinations between 45º and 30º also in the subclusters. The bottom panel adds galaxies deemed to be cluster members (or in the extended Virgo complex) from their morphology as listed in the last two sections of Table 3 of TFS. These are the outsiders in Table 2.

Fig. 4. Schematic of the Tully-Fisher correlation of line width and apparent magnitude for galaxies in a cluster that are all at the same distance. Shown is the incompleteness of the sampling in given line-width intervals for an apparent magnitude cutoff at B = 11.5. The right

hand line-width interval is unbiased because the complete line-width specific luminosity function is sampled. The smaller line widths (toward the left) are progressively biased at this magnitude-limit cutoff. The total incompleteness bias effect is calculated by summing over all line widths for magnitudes that are brighter than various observationally imposed magnitude cutoffs.

Fig. 5. The variation of the individual calculated moduli in the listed line-width intervals with apparent magnitude from the data in Table 2. The total sample (subclusters A plus B plus the "outsiders") is used. The correct mean modulus can be obtained only when each distribution is sampled completely. The completeness magnitude varies with line width, being complete if the sampling is made to B = 12 for the line-width interval with log LW > 2.5, but not complete for the smallest line-width interval of 2.2 to 2.1 until B = 15.

Fig. 6. The bias as a function of the cut-off magnitude for the complete Virgo cluster (total) sample from Table 2. The top panel shows the variation of <m - M> with depth of penetration into the cluster luminosity function. The bottom panel shows the variation of the calculated rms of the <m - M> as the magnitude cut off is varied. The intrinsic (true) rms value is not reached until B ~ 14. The data are from Table 4, cols. 7 and 8.

Fig. 7. The bias-free distribution of calculated distance moduli for the Virgo cluster sample from Table 2 using the calibration from equation (2). The subclusters A and B plus the "outsiders" are combined here in the top panel, and are shown separately in the lower two panels. Most of the spread is due to the intrinsic dispersion of the TF method (Fig. 2) rather than a real variation in the distances of galaxies in the sample.

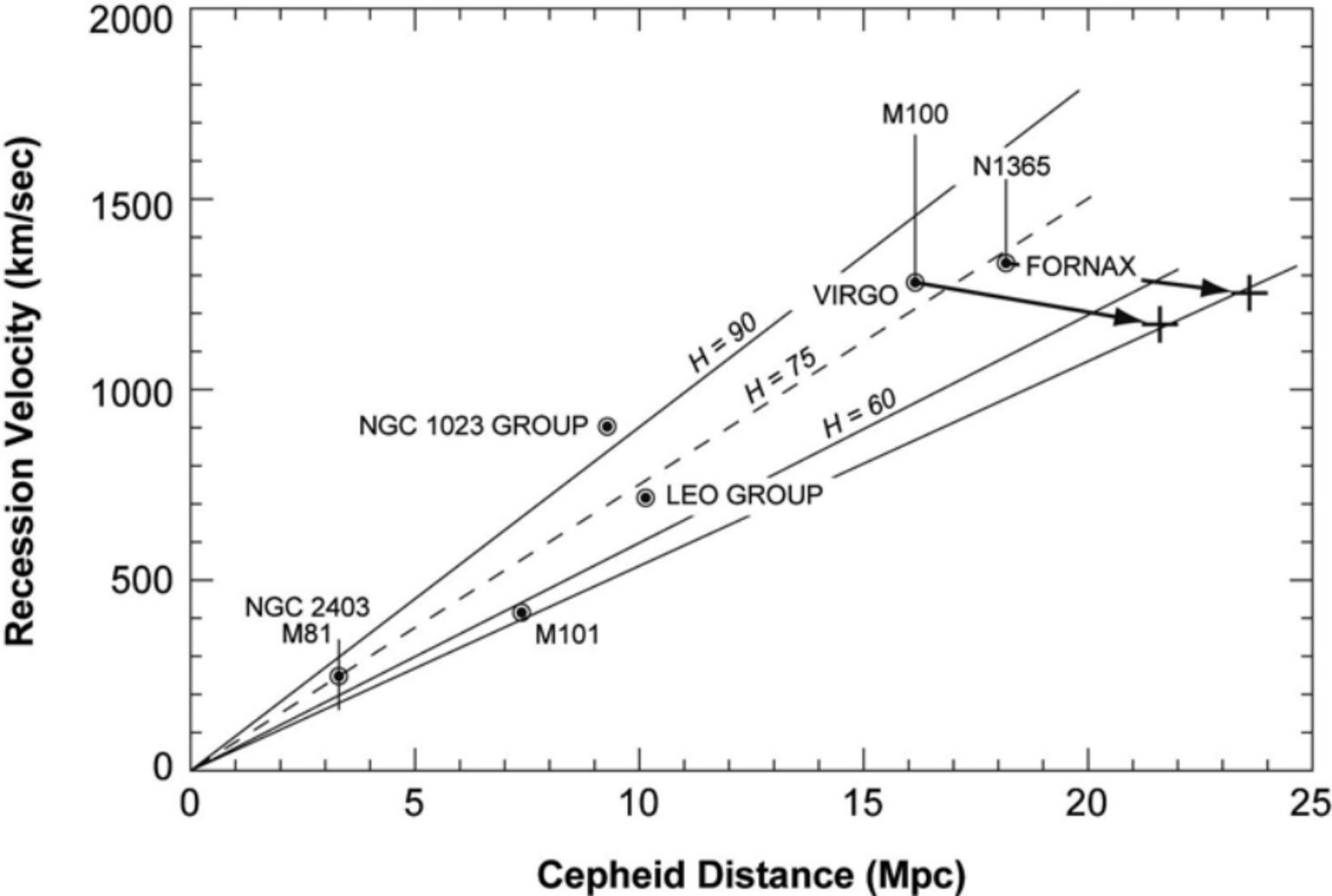

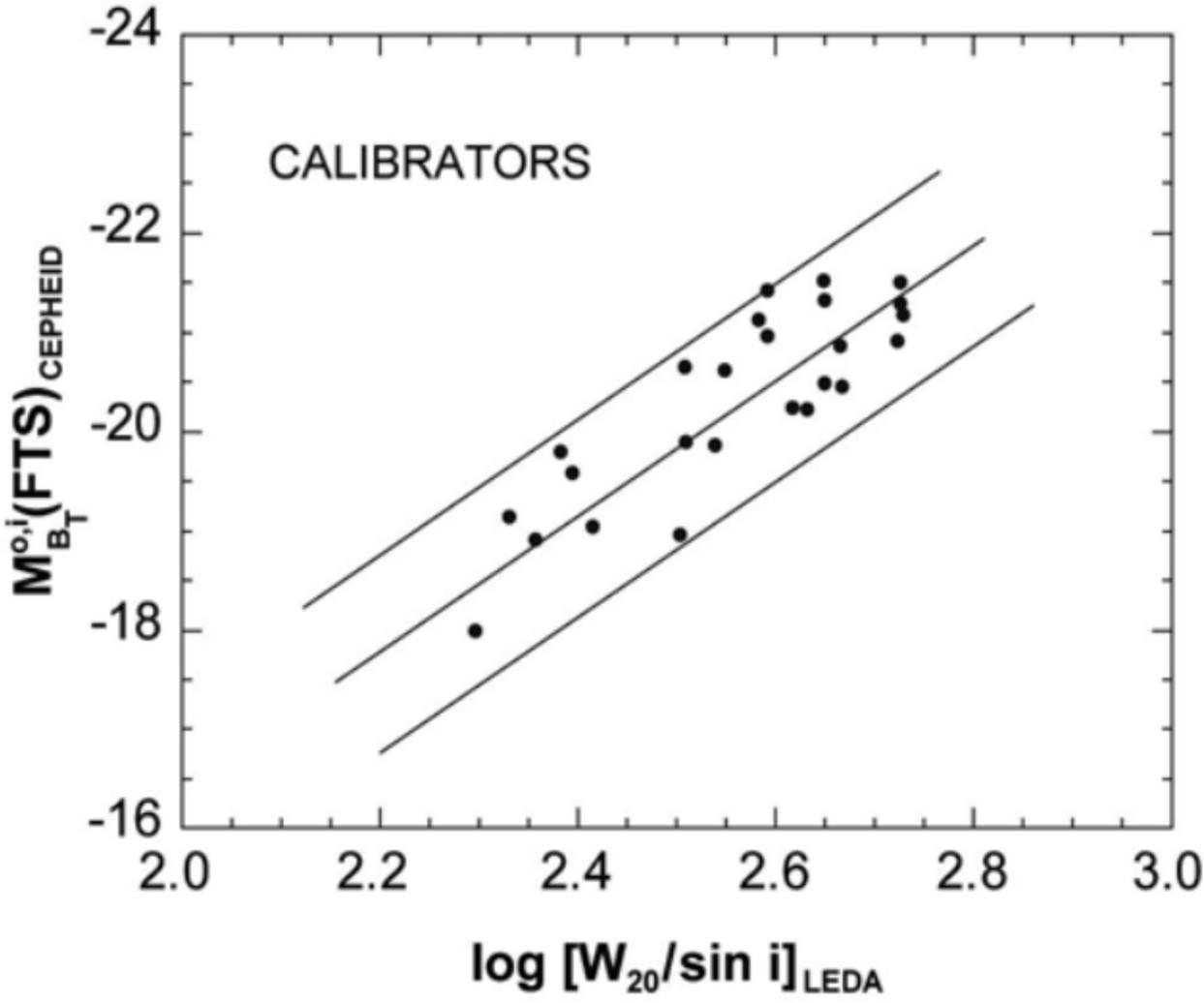

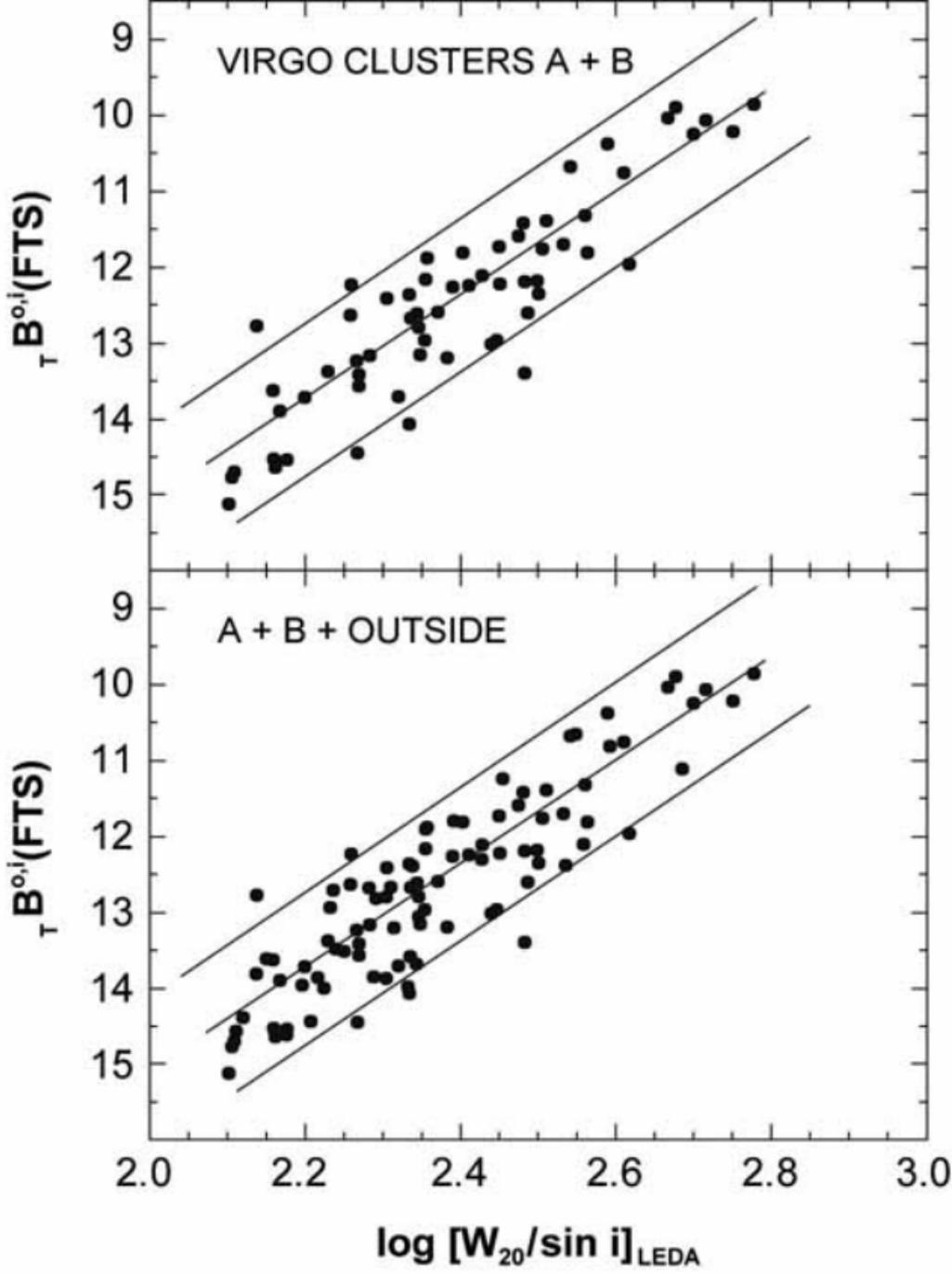

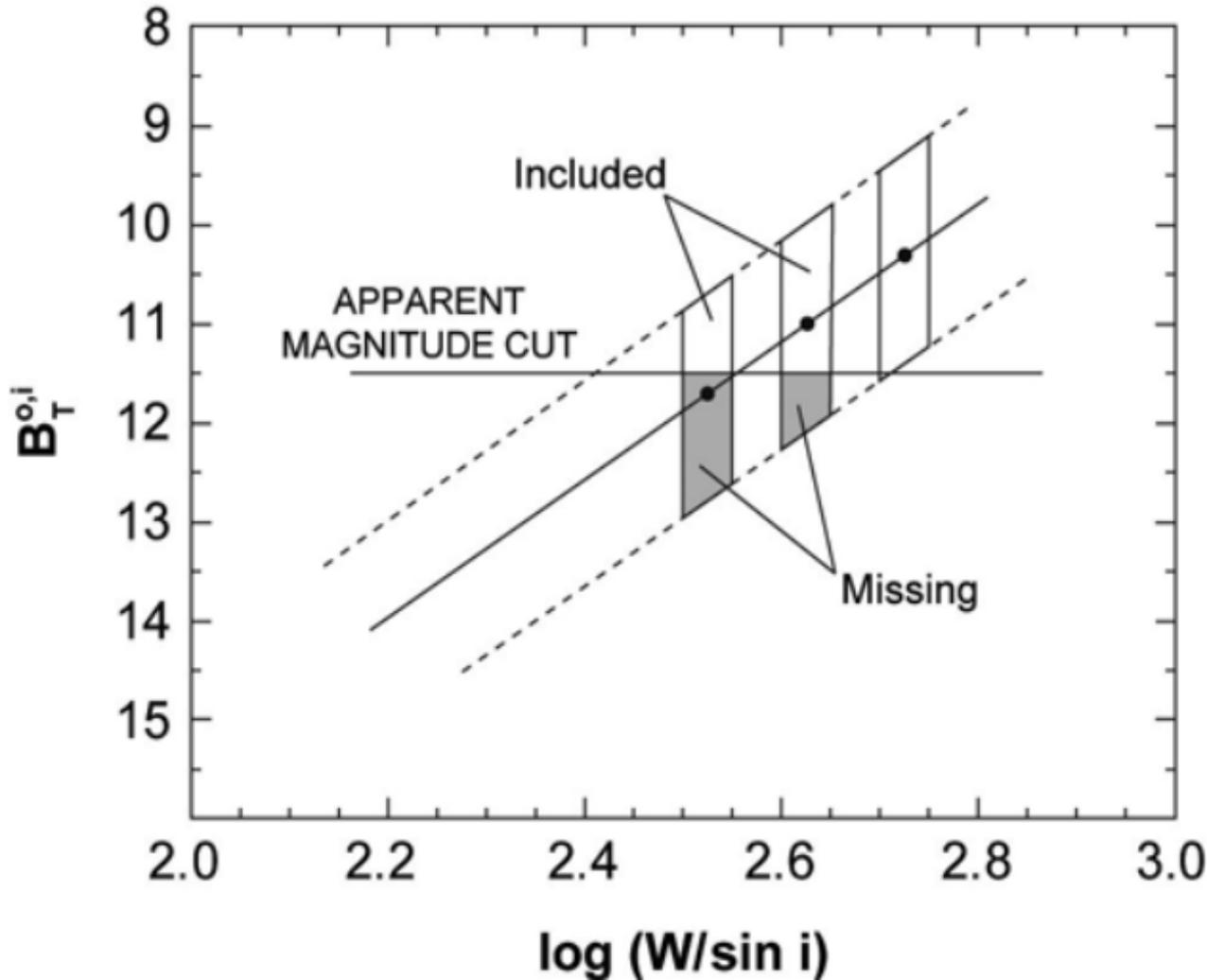

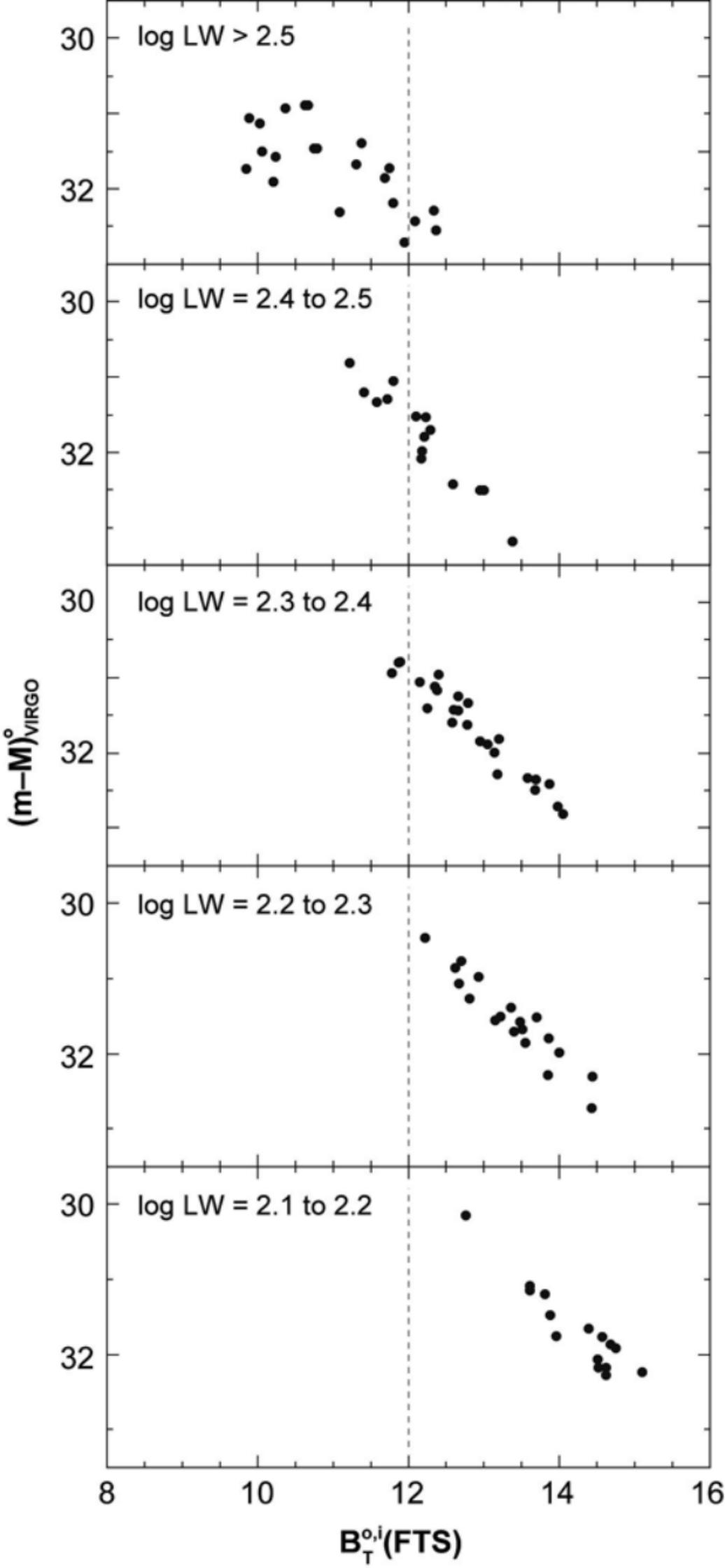

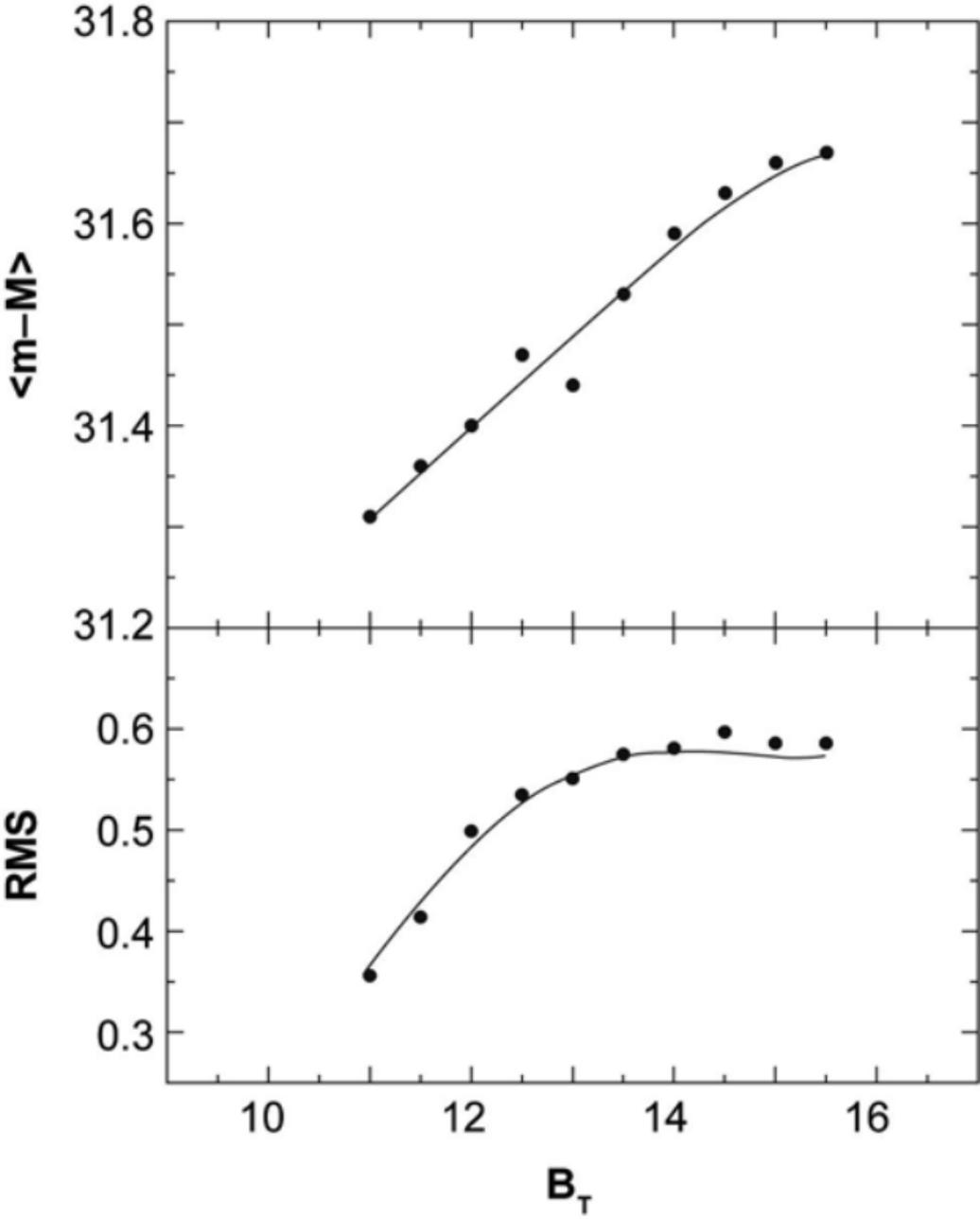

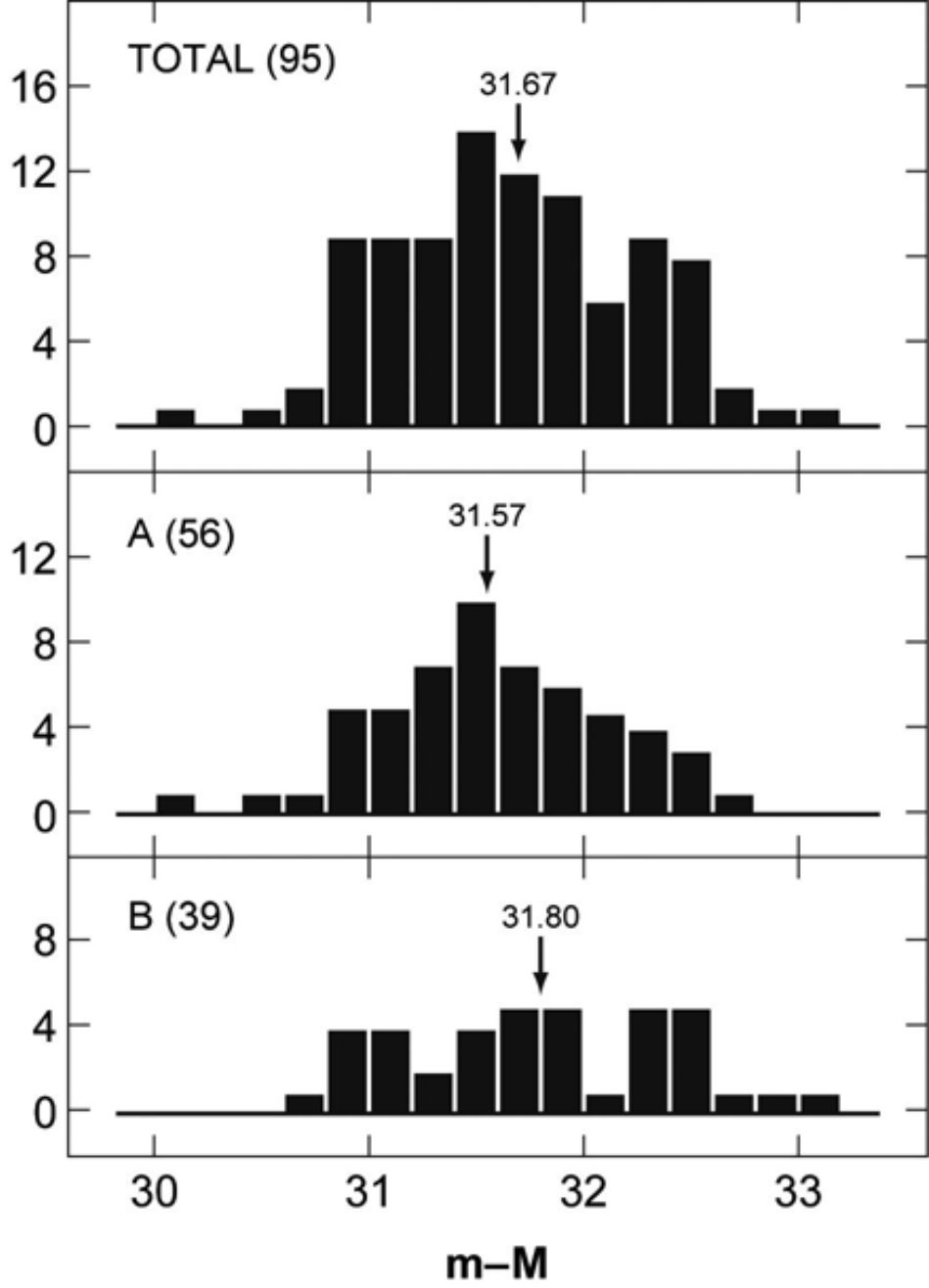